\documentclass[a4paper,twoside]{article}
\newcommand{\affil}[1]{$^{\rm #1}$}
\usepackage[authoryear]{natbib}
\bibpunct{(}{)}{;}{a}{}{,}
\usepackage{graphicx}
\date{} 
\title{\large\bf\flushleft The Warped Disk of Centaurus A from 
a radius of 2 to 6500 pc}
\author{\parbox{\textwidth}{\flushleft
\vspace{-0.5cm}
{\it Alice C. Quillen\affil{A}, Nadine Neumayer\affil{B}, Tom Oosterloo\affil{C}, and Daniel Espada\affil{D,E}}\\
\vspace{0.4cm}
{\small \affil{A}\,Dept. of Physics and Astronomy, University of Rochester, Rochester, NY 14627, USA; Email: aquillen@pas.rochester.edu} \\
{\small \affil{B}\,European Southern Observatory, Karl-Schwarzschild-Str 2, 85748 Garching, Germany}\\
{\small \affil{C}\,Netherlands Foundation for Research in Astronomy, Postbus 2, 7990 AA Dwingeloo, the Netherlands; Kapteyn Astronomical Institute, University of Groningen, PO Box 800, 9700 AV Groningen, the Netherlands}\\
{\small \affil{D}\,Harvard-Smithsonian Center for Astrophysics, 60 Garden St., Cambridge, MA 02138, USA }\\
{\small \affil{E}\,Instituto de Astrofisica de Andalucia (CSIC), Apdo. 3004, 18080 Granada, Spain}\\
}}
\begin{document}
\twocolumn[
\begin{changemargin}{.8cm}{.5cm}
\begin{minipage}{.9\textwidth}
\vspace{-1cm}
\maketitle
%
%
\small{\bf Abstract:}

We compile position and inclination angles for tilted
ring fits to the warped dusty and gaseous disk 
spanning radius 1.8 to 6500 pc from recent
observations.  For radii exterior to a kpc,
tilted ring orientations lie on an arc on a polar 
inclination versus position angle plot,
suggesting that precession following
a merger can account for the ring morphology.
Three kinks in the ring orientations 
are seen on the polar plot, the one at radius of about 1.3 kpc we suspect
corresponds to the location where self-gravity in the disk
affects the ring precession rate.
Another at a radius of about 600 pc 
may be associated with a gap in the gas distribution.
A third kink is seen at a radius of 100 pc. 
A constant inclination tilted disk precessing about the jet axis may
describe the disk between 100 and 20 pc but not interior to this.  
A model with
disk orientation matching the molecular circumnuclear disk at 100 pc
that decays at smaller radii to an inner flat disk perpendicular to the 
jet may account for disk orientations within 100 pc.  Neither model
would account for the cusps or changes in disk orientation at 100 or 600 pc. 


\medskip
\medskip
\end{minipage}
\end{changemargin}
]
\small

\section{Introduction}

In its central regions, NGC\,5128 exhibits a well recognized, optically-dark
band of absorption across its nucleus. This dusty disk was first modeled as a
transient warped structure by \citet{tubbs80}.
The kinematics of the ionized and molecular gas are well modeled by
a warped disk composed of a series of inclined  connected rings undergoing
nearly circular motion
\citep{bland86,bland87,nicholson92,quillenCO}.
Recent work has better fit the morphology of the disk in the infrared
using Spitzer Space Telescope observations \citep{quillen_irwarp}
(and improving on the model fit to disk seen in absorption
in the near-infrared; \citealt{quillen93}), 
with high resolution interferometric molecular CO observations and covering the
central arcminute \citep{espada09}, 
tracing the warm molecular hydrogen to within the sphere of
influence of the black hole using adaptive optics \citep{neumayer07,neumayer09},
and with new HI observations \citep{osterloo09}.

\begin{figure}[h]
\begin{center}
\includegraphics[scale=0.65, angle=0]{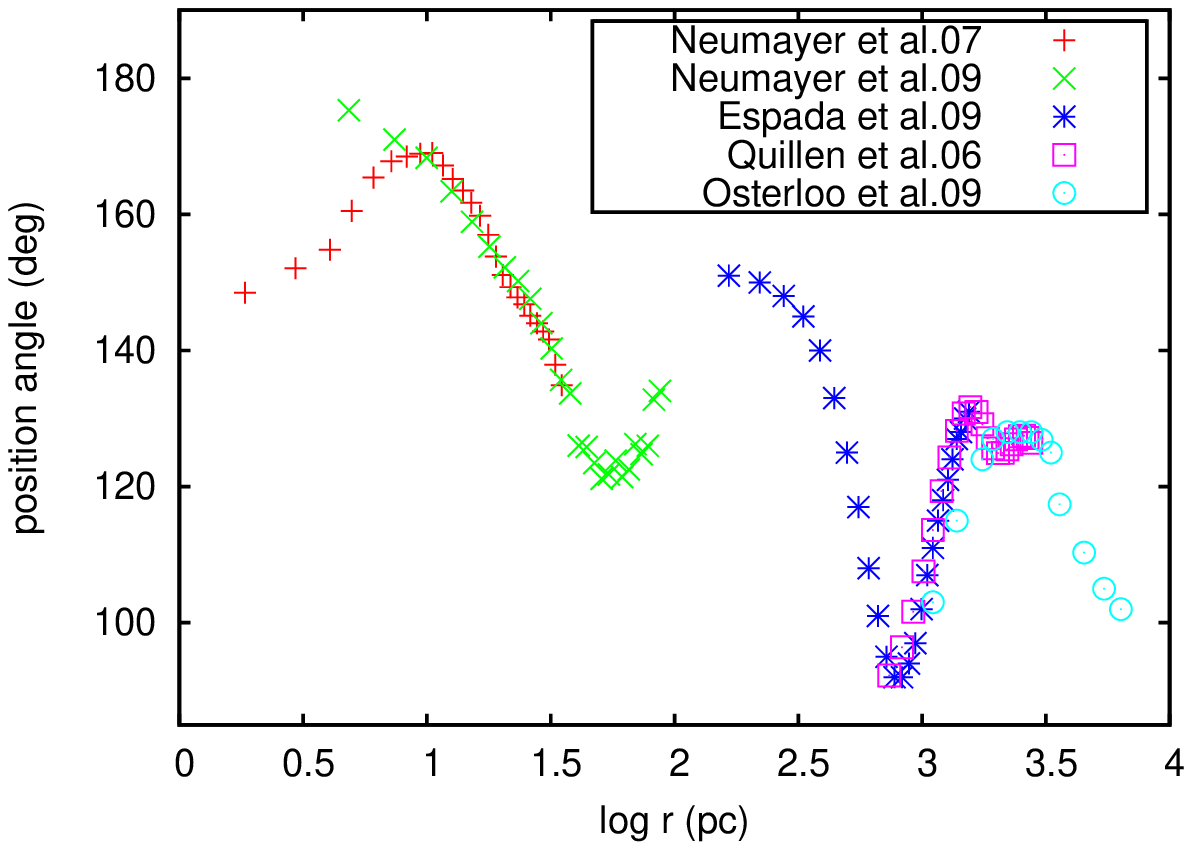}
\includegraphics[scale=0.65, angle=0]{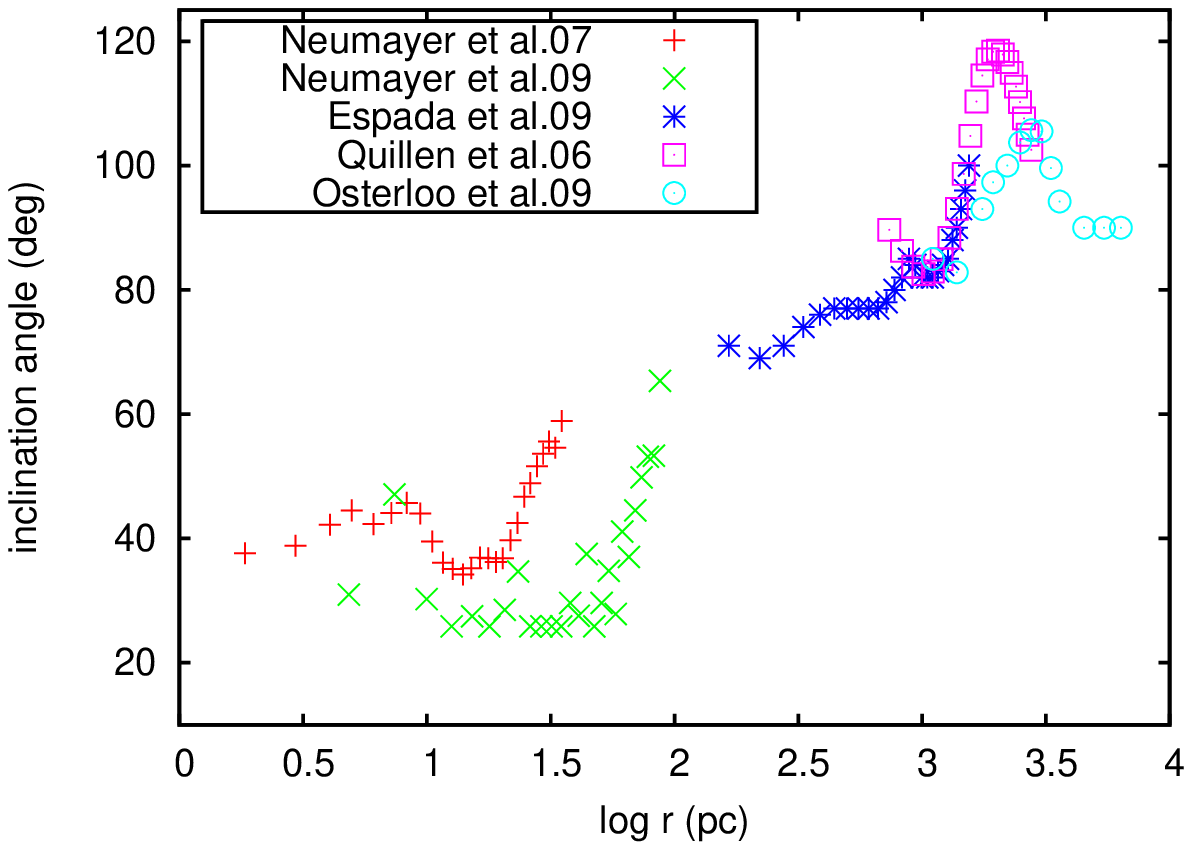}
\caption{Position and inclination angle as a function of radius
compiled from tables and figures by
\citet{neumayer07,neumayer09,espada09,quillen_irwarp,osterloo09}.
Data used to make these plots are listed in Table 1.
}
\label{fig:fig1}
\end{center}
\end{figure}

We take the opportunity to compile tilted
ring fits to the disk morphology.
The result is a warp geometry, described by tilted ring major axis 
position and inclination angles
as a function of radius, that spans from radius $0.1''$ to
350$^{\prime\prime}$ or using a distance of 3.8~Mpc  to Centaurus A \citep{harris09}
from 1.8~pc to 6.5~kpc. Here the smallest scales are set by
the $0.12''$ K band resolution of the adaptive optics
observations \citep{neumayer07} and the largest
scales by the extent of the tilted ring fit to the HI \citep{osterloo09}.
The factor of 3600 between the smallest and largest scales
is large compared to other studied gaseous disks.
This large scale of observations allows us to search
for orientation changes caused by different physical processes.
In the outer regions the primary processes causing orientation
changes are torques from the galactic potential and from 
perturbations caused by galaxy mergers. 
Interior to these torques may be caused by 
a variety of processes including
self-gravity in the disk \citep{sparke96}, radiation pressure 
(e.g., \citealt{pringle96,maloney96}), winds or pressure gradients in
the ambient ISM \citep{quillen99,quillen01}, 
and the Bardeen-Petterson effect (e.g., \citealt{caproni07}).

\section{Compiling Tilted Ring Fits}

Position and inclination angle for the tilted
ring fit to the warm molecular hydrogen emission 1-0~S(1) H$_2$ line
at 2.122$\mu$m were taken from Table 1 by \citet{neumayer07}
and have a spatial resolution of 0.12$^{\prime\prime}$, 
covering the central $3^{\prime\prime} \times 3^{\prime\prime}$.
The submillimeter SMA CO observations have a non-circular beam
with widest direction 6$^{\prime\prime}$ FWHM
and cover a 1$^\prime$ field of view  \citep{espada09}.
Angles used are those used to make Figure 8a by \citet{espada09} and are those
for the tilted ring fit lacking a bar perturbation.
The angles from the fit to the 8$\mu$m morphology seen in  Spitzer Space
Telescope imaging with the IRAC camera
are those used to make Figure 7 by \citet{quillen_irwarp}.
The IRAC camera at 8$\mu$m has a resolution of 2.2$^{\prime\prime}$.
The angles from the tilted ring
fit to the new HI observations are by \citet{osterloo09}.
There are no angles listed within 60$^{\prime\prime}$  
from the HI fit because HI is depleted
in the center of the galaxy.
The fifth set of position and inclination angles
are from observations and an associated tilted ring fit
that will be described in more detail by \citet{neumayer09}.
The measurements are from integral field spectroscopy with SINFONI 
and are taken in natural seeing under excellent conditions of 
0.5$^{\prime\prime}$, covering 
a 8$^{\prime\prime}\times 8^{\prime\prime}$ 
field of view (see \citealt{cappellari09}) 
and using the same 1-0~S(1) H$_2$ line discussed by \citet{neumayer07}.  
Tilted rings were fit to the velocity field 
using the procedures described by \citet{neumayer07}.
The compiled position and inclination angles from these
five studies as a function of
log radius in pc are shown in Figure \ref{fig:fig1}.
These angles are also listed in Table 1.  We note that two sets
of these tilted ring fits \citep{neumayer09,osterloo09}
are preliminary and may be updated in future.

\begin{figure}[h]
\begin{center}
\includegraphics[scale=0.9, angle=0]{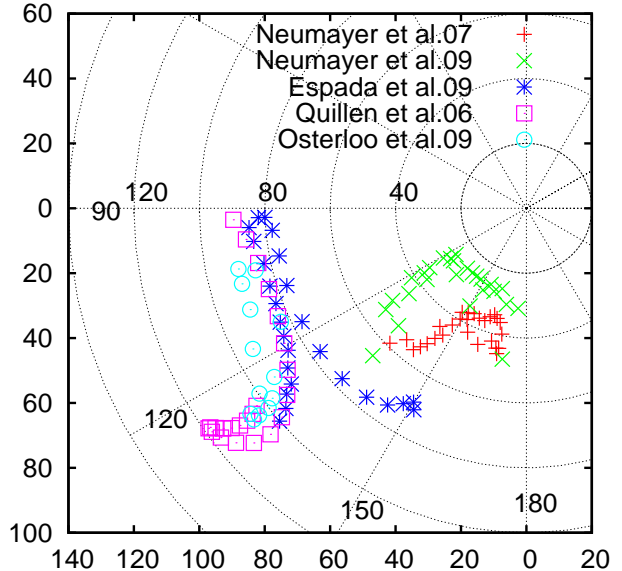}
\caption{Polar plot of ring position and inclination angles.
Here points are plotted with radius on the plot
given by the ring inclination and 
at an angle on the plot set by the ring major axis position angle.
The origin (with zero inclination) is on the upper right.
The same data are shown here as in
Figure \ref{fig:fig1}.   
We note that the disk near the nucleus lies at quite a different
orientation than the outer disk.
Three kinks or cusps are identified on this plot.  The first on
the lower left and in both the HI and Spitzer points 
(turquoise circles and pink squares).  
The second on the upper left and primarily in the CO points (blue stars).
The third between the CO and H$_2$ fits on the lower right (blue stars and
green X-s).
}
\label{fig:fig2}
\end{center}
\end{figure}

The inclination angle is defined such that an edge-on disk is oriented
with $i=90^\circ$.   Inclinations above $90^\circ$ refer to those with
northern side closer to the viewer (and so masking the northern
side of the galaxy) and those below $90^\circ$ refer to
the opposite.  The tilted ring fit to the disk seen
in extinction (e.g., \citealt{quillen93})
specify which side of the disk is closer to the viewer.
The position angle of a ring is that on the sky (counter clockwise from north)
of the blue shifted side of its major axis.

There is consistency between the different tilted
ring fits in the outer galaxy 
with the HI based kinematic fit at lower inclination
than that done from the Spitzer images. 
The fit to the Spitzer images was done without kinematic 
constraints.  Because the model was created with a 3D
data cube, it was computed faster when the cube was smaller.
This likely biased the best fit model to configurations more
twisted and compact than the real outer disk.   The HI kinematic fit
should be considered more accurate than the model
used to match the Spitzer images. 

The CO data appear to connect
the outer and inner regions though we must keep in mind
that the molecular gas distribution is dominated in the center
by a 100--200~pc radius circumnuclear disk 
and the gas distribution may contain one or more gaps.
The innermost points of the CO ring fit arise from the circumnuclear
disk itself and so have well measured position angles.
Consequently it is unlikely that the 
innermost position angles have been overestimated. 

The fits to the $H_2$ emission by \citet{neumayer07} and \citet{neumayer09}
overlap and have consistent position angles, 
but those from the smaller field of view data
have about 20$^\circ$ higher inclinations.
We expect that the fits to high resolution observations
by \citet{neumayer07}
would be more accurate in the center, 
and those to observations covering a larger field of view 
by \citet{neumayer09} would be more accurate
at larger radii.  However further analysis (and possibly deeper
observations) are required to resolve the discrepancy.

\begin{figure}[h]
\begin{center}
\includegraphics[scale=0.90, angle=0]{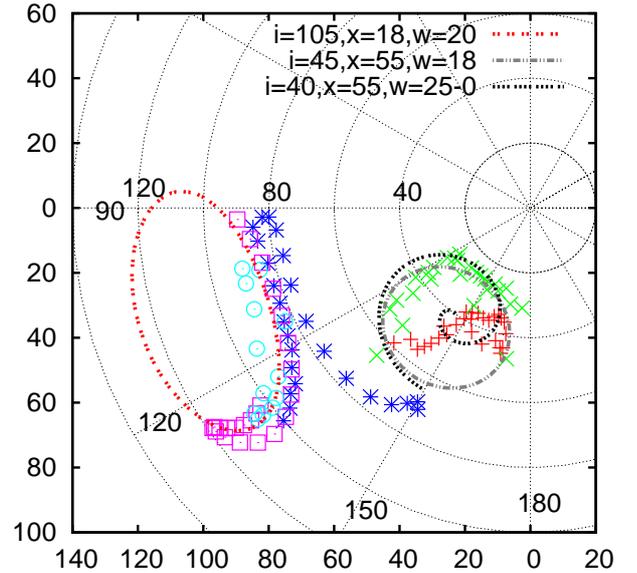}
\caption{Same as Figure \ref{fig:fig2} but with the addition
of two ovals and a black spiral. The ovals 
show orientations lying at fixed angular distance from 
a particular orientation on the sky.  The  spiral shows
a disk that approaches an orientation perpendicular to 
the jet axis at small radius.
The outer disk orientations lie on an arc which can be 
explained with precession following a merger. Precession occurs
about a symmetry axis that is estimated from the center
of the red oval (on the left).
The gray oval (on the right) shows ring orientations that have
angular momentum axis 18$^\circ$ from
orientation with inclination 45$^\circ$ and position
angle the same as the radio jet (55$^\circ$; \citealt{burns83}).
The black spiral shows a model with inclination and position
angle the same as the circumnuclear disk at 100 pc but that
approaches orientation perpendicular to the jet axis at small radii.
}
\label{fig:fig3}
\end{center}
\end{figure}

\subsection{Outer disk}

In Figure \ref{fig:fig2} we show the same angles listed
in Table 1 and shown in Figure \ref{fig:fig1} but 
on a polar plot.  Here the radius of points on the plot is set by
the inclination angle and the angle of points on the plot
set by the position angle of the tilted rings on the sky.
We first discuss  the disk exterior to 1~kpc.
We see that the ring angles from the outer galaxy lie on an arc.
Were we to plot the ring orientations on a sphere
the arc on this plot could correspond to part of a circle
centered about a particular orientation.
The rings in the inner galaxy lie at quite different orientations
compared to the rings in the outer galaxy.

In Figure \ref{fig:fig3} we show the same polar plot
 as in Figure \ref{fig:fig2} but with two separate ovals.
Each oval shows 
vectors that are the same angular distance from 
a particular orientation.    
The red oval (on the left) corresponds to all rings that have angular momentum
axis 20 degrees from a direction that is 105$^\circ$ inclined
from the line of sight and at a position angle of $18^\circ$ on the sky.
This is similar to the axis used to set the underlying galaxy symmetry axis
assumed by \citet{quillen93,quillen_irwarp} when generating precession
models for the disk,
though in neither model was the disk inclination held
fixed with respect to the galaxy axis.
The model by \citet{quillen93} has inclination angle with 
respect to the assumed underlying galaxy symmetry axis that increases
with increasing radius whereas that used to fit the Spitzer data
decreases with increasing radius.  

The arc in the polar plot from the outer disk can be 
accounted for with a merger model.
Following a merger, gas accreted from another galaxy would
have angular momentum that is not necessarily aligned
with the symmetry axis of the underlying galaxy.
In a coordinate system with respect to a galaxy symmetry
axis (assuming the galaxy is approximately axisymmetric)
each gas ring would maintain its inclination but
would precess about the galaxy symmetry axis.
The angular difference between the gas angular momentum and the galaxy
symmetry axis sets the size of the oval whereas the galaxy
symmetry axis orientation determines the center of the red oval (on the left)
in Figure \ref{fig:fig3}.

A simple precession model will not give a perfect fit or prediction
because mergers are not expected to leave
a gas remnant restricted to a single plane (corresponding
to a very narrow distribution of angular momentum orientation),
the underlying galaxy itself may be triaxial rather than
axisymmetric and 
would not be static during the merger.
The outer HI ring orientations ($r>1.6$~kpc or 85$^{\prime\prime}$) 
lie shifted but not
too distant from ring orientations in the disk between
0.8 --1.5 kpc (or 40 -- 80$^{\prime\prime}$), suggesting
that the angular momentum vector of gas stripped in the
outer regions was not too different from that forming the middle regions
of the disk.

Within about 1.3~kpc the position angle increases with increasing radius
rather than decreases with radius as it does exterior to 1.3 kpc 
(Figure \ref{fig:fig1}a).
This is also seen as a kink or cusp in the points on the lower 
left in the polar plot (Figure \ref{fig:fig2}).
A tilted ring 
precesses at a rate proportional to the ellipticity of the
gravitational potential and the angular rotation of
a particle in a circular orbit. Rotation rates
are faster in the center of the galaxy so a warped disk
should be increasingly corrugated in the center. 
A reduction in the rate or reversal of direction of precession can
cause a cusp in the orientation angles.
This kink would either be caused
by a variation in galaxy eccentricity, but could also be
due to self-gravity in the disk \citep{sparke96}.
The direction of precession in the outer galaxy and isophotal 
major axis at a PA of about $20^\circ$ suggests that
either the galaxy is prolate and the disk is oriented near in its symmetry
plane or the galaxy is oblate and the disk is in a nearly polar orbit
\citep{quillen93,sparke96}.
A sharp change in galaxy eccentricity is not seen in the $3 \mu$m isophotes
suggesting that future warp models should reconsider the importance
of self-gravity as well as updated mass models for the stellar mass
distribution in accounting for the disk morphology. 

The outer disk at 8 and 24~$\mu$m and in HI emission is not symmetrical about
the origin (e.g., see Figure 2 by \citealt{quillen_irwarp}) 
implying that the outer disk is somewhat lopsided.  The degree of lopsidedness
could be precisely measured from the HI kinematic observations.
By studying the evolution of this asymmetry it might be possible
to derive constraints on the time since the merger and on the nature of the
merging event itself.

\subsection{Inner Disk}

In the polar plot (Figure \ref{fig:fig2}) there are at least 3 locations where
there are abrupt changes in orientation.  One change occurs
at a radius of about 35$^{\prime\prime}$  or 600~pc, the radius at which
\citet{quillen_irwarp}  failed to find
a good model for the Spitzer images and so 
suggested that there might be a hole in the gas and dust distribution.
\citet{espada09} found evidence for a gap in the gas distribution between
a radius of 200 and 800~pc and also showed that the fits could
be improved with a model that included non-circular motions the gas.
In Figure \ref{fig:fig2} the radius of the gap in the gas distribution  
corresponds to the topmost plotted point where 
there is a cusp in the CO generated points.

There is a third cusp on the polar angle plot
in Figure \ref{fig:fig2}
on the lower right corresponding to orientation angle
of the circumnuclear molecular disk at a radius of 100--200 pc 
(6--12$^{\prime\prime}$).
Orientations based on 
the near-infrared spectroscopy by \citet{neumayer09} 
approach that of the circumnuclear molecular disk, 
suggesting that there is cusp in the angular orientations at
a radius of about 100 pc or 6$^{\prime\prime}$.

The CO circumnuclear molecular disk
could be smoothly connected to the disk seen in warm molecular hydrogen
within 100~pc.
Both the circumnuclear molecular disk at radius 100--200~pc
and the innermost rings at a radius of 2~pc 
are approximately perpendicular
to the jet axis, though the innermost points have lower
inclinations than the circumnuclear disk. 
The radio and X-ray jets 
have a position angle of $55^\circ$ \citep{burns83,kraft00}
with the northern jet suspected to be closer to the observer.
VLBI studies suggest that 
the jet axis lies between 50 and 80$^\circ$
from our line of sight \citep{tingay98}, however apparent motions 
suggest an angle between 20 and 40$^\circ$ \citep{hardcastle03}. 
A disk perpendicular to the jet would have a major axis position angle
of 145$^\circ$.
If the disk at the smallest measured
radii is perpendicular to the jet then the jet is
about 40$^\circ$ from the line of sight.

On Figure \ref{fig:fig2} the points by \citet{neumayer09}
(green X-s)
are at higher inclination than those by \citet{neumayer07} (red plus signs).
Both fits at $r\sim 80$~pc begin near the orientation
of the circumnuclear disk on the left in Figure \ref{fig:fig2} and
Figure \ref{fig:fig3}. With decreasing
radius, the green points (X-s) \citep{neumayer09} move upwards (to
lower inclination) and then
to the right, whereas the red points (plus signs) \citep{neumayer07}
move to the right and then back to the left at the smallest radii.

Perhaps the circumnuclear molecular disk orientation and that
seen interior to it can be connected with an arc
corresponding to precession about an axis near the radio jet.
Orientations $18^\circ$ from an angular momentum axis
inclined 45$^\circ$ from the line of sight and at a position
angle of 55$^\circ$ are shown as a gray oval (on the right) 
in Figure \ref{fig:fig3}.
The gray oval passes near most
of the points by \citet{neumayer09} but is inconsistent with
the orientation in the inner disk \citep{neumayer07}.
The jet inclination angle is uncertain 
(see discussion by \citealt{hardcastle03}),   
but even if we changed the estimated jet inclination angle
and so moved the center of the oval, it 
would not pass through all the data points interior to
the circumnuclear molecular disk in Figure \ref{fig:fig3},
particularly those at radii within 10 pc by \citet{neumayer07}.

We have also plotted on Figure \ref{fig:fig3} a black spiral
that has an orientation near the circumnuclear disk
and approaches an orientation perpendicular to the jet axis at 
a radius of 1~pc.
For this model, the angular distance between the angular
momentum of the circumnuclear disk
at 100~pc and the
jet axis is 25$^\circ$.  The angle between
the angular momentum vector and jet axis smoothly drops to zero
with a slope $d i/d\alpha = -0.05$ where $i$ is inclination
and $\alpha$ precession angle with respect to the jet axis.
as the precession angle advances by 450$^\circ$. This
spiral was projected assuming that
jet axis is $40^\circ$ from the line of sight.
This model passes near data points at radii interior to 10 pc,
suggesting that simultaneous precession and damping
to a plane perpendicular to the jet axis may be occurring in the disk
at radii interior to 100 pc.

The gravitational field of the black hole 
dominates that from the stellar component within
a radius known as the sphere of influence, 
$r_s = GM_{BH}/\sigma^2$. For a black hole mass of 
$M_{BH} = 5.5 \pm 3.0 \times 10^7 M_\odot$  
and velocity dispersion of about 150 km/s \citep{cappellari09}
this radius is about $r_s\sim 10$ pc, consequently the innermost
data points by \citet{neumayer07} lie within the massive black hole's
sphere of influence.
For comparison, the gravitational radius of the black hole at the nucleus is 
$r_g = 2GM_{BH}/c^2 = 5 \times 10^{-6}$~pc.
Relativistic frame dragging near $r_g$ for a spinning black hole 
leads to precession known as Lense-Thirring precession and
warping of a disk that is initially misaligned with the black hole
spin axis.  This is called the Bardeen-Petterson effect.
Due to viscous processes, the disk could settle into the plane
perpendicular to the black hole's spin axis as it accretes.  
The transition or Bardeen-Petterson 
radius is estimated to be of order $10^4 r_g$ 
\citep{caproni07}
corresponding to 0.1 pc for Centaurus A.  
This is well within that resolved by \citet{neumayer07}.
A similar situation exists for NGC 4258 with only the inner radius
of the resolved warped disk near the estimated Bardeen-Petterson radius
\citep{caproni07}.
The discrepancy between the estimated
Bardeen-Petterson radius and the observed location of the warp
might be resolved if there is a smooth transition between the
outer disk and the Bardeen-Petterson radius, as exhibited
by simulations of diffusive disks (e.g., \citealt{lodato06}).
The warped disk of NGC 4258 has been interpreted in terms of the 
Bardeen-Petterson
effect even though the warped disk itself lies outside the estimated
Bardeen-Petterson radius \citep{caproni07}.
The rough correspondence between the black spiral and data
points in the inner disk in Figure \ref{fig:fig3} suggest that
a similar model might be applied to Centaurus A's disk within
a radius of 10 pc.  Such a model would be a better fit for a jet
axis near 40$^\circ$ from the light of sight, 
consistent with the range estimated by \citet{hardcastle03}, 
but 10$^\circ$ lower than estimated from VLBI observations \citep{tingay98}.

Outside the sphere of influence of the black hole, $r_s$,
disk settling or warp diffusion timescales exceed the 
accretion timescale \citep{steiman88}.
However inside the black hole's 
sphere of influence the rotation is nearly Keplerian.
The warp diffusion timescale is significantly shorter for
Keplerian disks than for galactic disks \citep{papa83}.
The warp diffusion timescale for a Keplerian disk
$t_d \Omega \sim \alpha \left({h \over r}\right)^{-2}$,
\citep{papa83} where $h/r$ is the disk aspect 
ratio and $\alpha$ is the unitless parameter used to characterize 
the disk viscosity.
The Bardeen-Peterson radius (at which a disk would settle
to the midplane defined by the black hole spin axis) is estimated
by equating $t_d$ to the inverse of the Lense-Thirring precession rate.
The ratio of the Lense-Thirring precession rate to the angular
rotation rate is at most $10^{-8}$ at 1 pc from the black hole.
The expression for $t_d$ illustrates that
the Bardeen-Petterson radius could only be located near 1 pc if the disk
is extremely thin and viscous, so much so that it is uncomfortable to apply
such a model to the inner region of Cen A's disk.  
One could consider the possibility that
the disk does settle to a plane perpendicular to the jet axis
and it does so near and within $r_s$ where the warp diffusion
timescale is shorter than exterior to this radius.
If this is the case then there are other torques acting
on the disk that depend on orientation angle, for example
due to radiation pressure or pressure gradients
that cause precession faster than Lense-Thirring precession near $r_s$. 


A model with a smoothly changing orientation angle that decays
to an axis perpendicular to the jet would
not explain the orientations at radii 
between 100 and 600 pc, where there may be a gap in the gas distribution. 
Neither would a Bardeen-Petterson effect model
account for the cusps or changes in orientation at a radius of 100 pc
and at 600 pc.  One possibility
is that a triaxial bulge induces non-circular motions in this region 
complicating the measurement of orientations \citep{espada09}.
Another possibility is that bursts of star formation or nuclear activity
have blown out material from this region (e.g., \citealt{quillen_supershell}).
In this case, the kinks in the orientation
direction may be related to past episodes of activity.
This last possibility suggests that future work could
explore a relation
between orientation and episodic nuclear activity 
even at radii interior to 100~pc.
Orientation changes related to epochs of nuclear activity would be predicted
if either radiative or wind driven warp instabilities operate or
if there are variable gradients in the ambient medium associated
with jet propagation.

\section{Discussion}

Surprisingly a comparison of the tilted ring fits
shows at least three kinks in the orientation angles
as a function of radius on a polar plot.
The first (at a radius about 1.3~kpc or 70$^{\prime\prime}$)
was previously known \citep{quillen93} and could correspond
to a variation in precession rate caused by self-gravity of the disk.
The outer disk forms an arc on a polar plot of ring orientation
angles implying that precession following a merger, as proposed
by \citet{tubbs80}, is a good explanation for the disk morphology.

Between the hole in the dust distribution at a radius
of about 30$^{\prime\prime}$  (or 600~pc) and the circumnuclear molecular
disk there is also an abrupt change in angle.
There is another change in orientation at a radius of about 100 pc 
(or 6$^{\prime\prime}$).
Precession about the jet axis 
could match the orientations between 100 and 20 pc,
but would not match the orientations interior to this. 
A model with disk angular momentum axis approaching the jet axis 
with decreasing radius could be promising. 
Models of the Bardeen-Petterson effect that are used to interpret
NGC 4258's disk might under extreme conditions 
be applied to Centaurus A even though
the resolved region (outside 1.8 pc) 
lies outside the estimated Bardeen-Petterson radius.
However, the two inner cusps or changes in the disk orientation would not
be predicted by a Bardeen-Petterson type of model.
Future work could explore models relating
disk orientation to past episodes of nuclear activity.

\section*{Acknowledgments} 
We thank Joss Bland-Hawthorn, Aneta Siemiginowska,  Richard Nelson
and Gordon Ogilvie for helpful discussions.
DE is supported by a Marie Curie International Fellowship (MOIF-CT-2006-40298) granted by the European Commission.
NN is supported by the DFG cluster of excellence 
`Origin and Structure of the Universe.'
Support for this work was in part provided by
by NASA through an award issued by JPL/Caltech, and
HST-AR-10972 to the Space Telescope Science Institute.
ACQ thanks the Newton Institute for support and 
hospitality during the fall of 2009.

\begin{table}[h]
\begin{center}
\caption{Inclination and Position Angles from Tilted Ring Fits}\label{tab:tab1}
\begin{tabular}{lcc|lcc|lcc}
\hline 
$r$ & Inc & PA  & $r$ & Inc & PA & $r$ & Inc & PA  \\
 ($^{\prime\prime}$) &(deg)&(deg)&($^{\prime\prime}$)&(deg)&(deg) &($^{\prime\prime}$)&(deg)&(deg) \\
\hline 
\multicolumn{3}{c|}{NEU07$^a$} & \multicolumn{3}{c|}{NEU09$^b$}& \multicolumn{3}{c}{ESP09$^c$} \\
\hline 
0.05 &   45.0  &    144.0  &   0.26  &   31.0 & 175.3  &  9  &    71 &  151 \\
0.10 &   37.6  &    148.5  &   0.40  &   47.1 & 171.0  &  12 &    69 &  150 \\
0.16 &   38.8  &    152.1  &   0.54  &   30.2 & 168.3  &  15 &    71 &  148 \\
0.22 &   42.2  &    154.8  &   0.68  &   25.8 & 163.4  &  18 &    74 &  145 \\  
0.27 &   44.5  &    160.5  &   0.83  &   27.4 & 158.9  &  21 &    76 &  140 \\  
0.33 &   42.3  &    165.4  &   0.97  &   25.8 & 155.5  &  24 &    77 &  133 \\  
0.39 &   44.1  &    167.8  &   1.12  &   28.5 & 152.2  &  27 &    77 &  125 \\
0.45 &   45.7  &    168.5  &   1.27  &   34.7 & 150.2  &  30 &    77 &  117 \\  
0.51 &   44.0  &    168.9  &   1.42  &   25.8 & 147.6  &  33 &    77 &  108 \\  
0.57 &   39.5  &    169.0  &   1.57  &   25.8 & 144.0  &  36 &    77 &  101 \\ 
0.63 &   36.1  &    167.2  &   1.73  &   25.8 & 140.3  &  39 &    78 &  95  \\ 
0.69 &   35.1  &    165.2  &   1.90  &   25.8 & 135.7  &  42 &    80 &  92  \\  
0.76 &   34.2  &    163.5  &   2.06  &   29.6 & 133.7  &  45 &    82 &  92  \\  
0.82 &   35.2  &    161.7  &   2.22  &   27.6 & 126.0  &  48 &    85 &  94  \\  
0.89 &   36.9  &    159.8  &   2.40  &   37.5 & 125.8  &  51 &    84 &  97  \\  
0.96 &   36.8  &    157.0  &   2.57  &   25.8 & 123.5  &  54 &    82 &  102 \\ 
1.03 &   36.2  &    153.8  &   2.75  &   29.6 & 121.1  &  57 &    82 &  107 \\  
1.10 &   36.8  &    151.1  &   2.94  &   34.8 & 121.8  &  60 &    82 &  111 \\ 
1.18 &   39.7  &    149.3  &   3.14  &   27.8 & 123.8  &  63 &    83 &  115 \\
1.26 &   42.5  &    147.8  &   3.34  &   41.1 & 121.4  &  66 &    84 &  118 \\  
1.34 &   46.7  &    146.8  &   3.55  &   37.0 & 122.5  &  69 &    85 &  121 \\ 
1.42 &   48.9  &    145.1  &   3.77  &   44.5 & 126.2  &  72 &    88 &  124 \\  
1.51 &   51.6  &    144.0  &   3.99  &   49.8 & 124.8  &  75 &    90 &  127 \\ 
1.60 &   53.6  &    142.8  &   4.23  &   53.1 & 126.0  &  78 &    93 &  128 \\ 
1.69 &   55.6  &    141.6  &   4.48  &   53.3 & 132.8  &  81 &    96 &  130 \\ 
1.79 &   54.6  &    137.9  &   4.74  &   65.3 & 134.1  &  84 &   100 &  131 \\ 
1.90 &   58.9  &    134.9  &         &        &        &     &       &      \\ 
\hline 
\multicolumn{3}{c|}{QUI06$^d$} &\multicolumn{3}{c|}{QUI06$^d$} & \multicolumn{3}{c}{OST09$^e$} \\
\hline 
 40  &   89.7  &   92   & 105 &   118.3 &   126  &  60    &  85.0  &   103.0 \\
 45  &   86.3  &   96   & 110 &   118.5 &   125  &  75    &  82.8  &   115.0 \\
 50  &   83.7  &   102  & 115 &   117.9 &   125  &  95    &  93.0  &   124.0 \\ 
 55  &   82.5  &   108  & 120 &   116.7 &   125  & 105    &  97.3  &   127.0 \\
 60  &   82.9  &   114  & 125 &   114.9 &   126  & 120    & 100.0  &   128.0 \\
 65  &   84.9  &   119  & 130 &   112.7 &   127  & 135    & 103.7  &   128.0 \\
 70  &   88.4  &   124  & 135 &   110.2 &   127  & 150    & 105.7  &   128.0 \\
 75  &   93.0  &   128  & 140 &   107.6 &   127  & 165    & 105.5  &   126.9 \\
 80  &   98.7  &   131  & 145 &   105.0 &   127  & 180    &  99.6  &   125.0 \\
 85  &   104.8 &   132  & 150 &   102.6 &   126  & 195    &  94.2  &   117.4 \\
 90  &   110.3 &   131  &     &         &        & 245    &  90.0  &   110.3 \\
 95  &   114.5 &   129  &     &         &        & 295    &  90.0  &   105.0 \\
 100 &   117.1 &   127  &     &         &        & 345    &  90.0  &   102.0 \\
\hline
\end{tabular}
\medskip\\
$^a$Angles are from Table 1 by \citet{neumayer07}.\\
$^b$Preliminary fits by \citet{neumayer09}.\\
$^c$Angles are from Figure 8a by \citet{espada09}.\\
$^d$Angles are from Figure 7 by \citet{quillen_irwarp}.\\
$^e$Preliminary fits by \citet{osterloo09}.\\ 
\end{center}
\end{table}

\end{document}